\documentstyle[psfig]{mn}

\ifnfsstwo

\fi

\ifnfssone
  \newmathalphabet{\mathit}
    \addtoversion{normal}{\mathit}{cmr}{m}{it}
    \addtoversion{bold}{\mathit}{cmr}{bx}{it}

\fi

\ifoldfss

\fi

  \makeatletter
  \ifx\CUP@mtlplain@loaded\undefined  
  \makeatother  
    
  \else
    
  \fi

\loadboldmathitalic 
\loadboldgreek      
 
\def\etal{{\rm et al.~}}
\def\simless{\mathbin{\lower 3pt\hbox
   {$\rlap{\raise 5pt\hbox{$\char'074$}}\mathchar"7218$}}}   
\def\simgreat{\mathbin{\lower 3pt\hbox
   {$\rlap{\raise 5pt\hbox{$\char'076$}}\mathchar"7218$}}}   
\def\etal{{\rm et al.}}

\def\solmas{{M$_\odot$}}
\def\solm{{M_\odot}}
\def\solrad{{R$_\odot$}}
\def\solr{{R_\odot}}

\def\trel {t$_{\rm relax}$}

\def\vdisp {$v_{\rm disp}$}
\def\vdsp {v_{\rm disp}}
\def\venc {v_{\rm enc}}

\def\tcol {t_{\rm enc}}

\def\rc {R_{\rm enc}}
\def\rcoll {R$_{\rm enc}$}
\def\rhd {R_{\rm hard}}

\def\ms {M_{\star}}
\def \pc3 {pc$^{-3}$}
\def \kms {km s$^{-1}$}
\def \spcc {stars pc$^{-3}$}
\def \au {\sc au\rm}
\def \aus {\sc au \rm}

\makeatletter
\ifx\CUP@mtlplain@loaded\undefined  
  \newfont\bit{cmbxti10 at 9pt}
\else
  \newfont\bit{mtbxti10 at 9pt}
\fi
\makeatother  
 
\title[Planets in stellar clusters] {Planetary dynamics in stellar clusters} \author[Bonnell et. al.]
{Ian A. Bonnell$^1$, Kester W. Smith$^{2,3}$, Melvyn B. Davies$^4$ and Keith Horne$^1$ \\ 
$^1$ School
of Physics and Astronomy, University of St Andrews, North Haugh, St
Andrews, Fife, KY16 9SS. \\
$^{2}$ Institut f\"ur Astronomie, ETH-Zentrum, CH-8092 Z\"urich, Switzerland. \\ 
$^{3}$ Paul Scherrer Institut, W\"urenlingen und Villigen, CH-5232 Villigen PSI,
Switzerland \\
$^{4}$ Department of Physics and Astronomy, Univeristy of Leicester, 
Leicester, LE1 7RH. \\ 
}
\pagerange{\pageref{firstpage}--\pageref{lastpage}}
\pubyear{    }

\def\LaTeX{L\kern-.36em\raise.3ex\hbox{a}\kern-.15em
    T\kern-.1667em\lower.7ex\hbox{E}\kern-.125emX}

\begin{document}

\label{firstpage}

\maketitle

\begin{abstract}

We investigate how the formation and evolution of extrasolar planetary
systems can be affected by stellar encounters that occur in the
crowded conditions of a stellar cluster. Using plausible estimates of
cluster evolution, we show how planet formation may be supressed in
globular clusters while  planets wider than $\simgreat 0.1$ \aus 
that do form in such
environments can be ejected from their stellar system. Less crowded
systems such as open clusters have a much reduced effect on any
planetary system.  Planet formation is unaffected in open clusters and
only the wider planetary systems will be disrupted during the
cluster's lifetime.  The potential for free-floating planets in these
environments is also discussed.

\end{abstract}


\section{Introduction}

The recent discovery of significant numbers of extrasolar planets 
(Mayor \& Queloz 1995, Marcy \& Butler~1996; Marcy~1999) has
driven an outbreak in research into planet formation. Previously, our
knowledge has been based entirely on one data point: our solar
system. 

The large increase in the number of known systems has had two major
consequences for our understanding of planetary formation and
evolution: firstly, it seems that planetary systems are not rare, and
secondly that they need not conform to solar system type
configurations. Specifically, the fact that the extrasolar planets
discovered so far are gas giants commonly in close orbits was
unexpected according to theories based upon the planets in the solar
system (eg Lissauer~1993; Ruden~1999).  
This has led to new theories to explain how
gas giants that form at distances similar to Jupiter from their
central star can migrate inwards to occupy the close orbits as has
been found (Lin, Bodenheimer \& Richardson~1996). Forming the gas
giants at such distances is seen as improbable due to the lack of 
sufficient condensable material for planetesimal growth.

A further complication to planet formation may arise due to the fact
that stars are commonly found and perhaps generally formed in stellar
clusters.  In addition to the well known globular and open clusters,
recent IR surveys have shown that most young stars are found in dense
embedded clusters (cf. Clarke, Bonnell \& Hillenbrand~2000).  The
density of these clusters range from $10^3$ to $\simgreat 10^4$ stars
pc$^{-3}$ in the core of the Orion Nebular Cluster (ONC). Larger
clusters such as R136 in 30 Doradus have even higher densities
($\simgreat 10^5$) and are probably more appropriate for the early
conditions of globular clusters. It is the aim of this paper to
investigate how the high stellar density in such regions affect
both planet formation and planetary survival. Cluster membership
has other disadvantages as the proximity of massive stars can
also act to impede planet formation (Armitage~2000).

\section{Cluster Evolution}

In determining how relevant stellar interactions are for planets and
planet formation, we have to consider not only the present cluster
conditions but also the cluster's previous evolution. Although it is
difficult if not impossible to determine the previous evolution in
individual cases, we can estimate probable evolutionary histories by
considering cluster dynamics and initial conditions. Firstly, if we
consider the young (embedded) clusters found in star forming regions,
they generally contain significant amounts of mass in the form of
gas. For example, the ONC is believed to contain 50 per cent of its
mass in gas (Hillenbrand \& Hartmann~1998). As the IMF and median mass
are typical of field stars (Hillenbrand~1997), the majority of this
mass will be ejected from the system. In general, the mass that is not
accreted will help unbind the cluster. The evolution of the cluster
undergoing gas removal depends critically on the gas fraction and on
the removal timescale (Lada, Dearborn \& Margulis 1984, Goodwin~1997).
Simulations of cluster expansion due to gas expulsion have shown that
for clusters that do survive, they generally increase their half-mass 
radii
by factors of $\approx 3-5$ (Kroupa, Aarseth and Hurley~2000;
Goodwin, Pearce \& Thomas~2000).  This corresponds to a decrease in the
mean cluster density of 10 to 100.  Another way of quantifying this
expansion is by comparing the youngest embedded clusters with older
open clusters. Thus, from densities of $10^{3}$ to $\simgreat 10^{4}$ \spcc\ 
typical of the ONC the clusters must evolve towards densities of
$\simless 10^{2}$ \spcc\ typical of open clusters. This is probably a lower
limit as some gas removal may have already occurred in the youngest
systems. Furthermore, the lower frequency of open compared to embedded
clusters implies that many of the embedded clusters do not survive the
expansion phase.  The timescale for this high density phase is likely
to be the lifetime of the most massive stars in the system. Especially
in systems with high velocity dispersions such as young globulars,
multiple supernova events are the likely cause of gas removal and thus
setting a timescale of several $\times 10^6$ years.

After gas removal, the cluster will continue to expand due to a 
combination of mass segregation, tidal interactions with the Galaxy 
which removes stars from the cluster (eg Terlevich~1987) 
and binaries which absorb a
large fraction of the binding energy. The combination of all these
effects can typically increase the cluster's radius substantially 
and thus decreases the stellar density. Simulations of
post core-collapse clusters including the effect
of binaries have shown a decrease in density of $\approx 10$ (Giersz 
\& Heggie~1997) when not filling their tidal radius.

From these considerations, we can estimate that cluster densities
generally decrease by factors of 10 to $\simgreat 100$ (or possibly up to 1000 in some cases) over their
evolution.  Although these higher density phases may have occurred over
much shorter timescales, they could potentially have had dramatic
effects on any planetary systems.

\section{Cluster properties}

The clusters we consider here range from the low density open clusters
in the stellar neighbourhood to the globular clusters such as 47 Tuc
and the possible precursers of both of these types of systems. Open
clusters generally have stellar densities of $\simless 10^2$ \spcc\
and ages of $\simless 10^9$ years with low velocity dispersions,
$\vdsp \simless 1$ \kms. Based on the above discussion, precursers of
such systems probably had densities of $\simgreat 10^3$ to $10^4$
(possibly as high as $10^5$) \spcc\ with still low velocity
dispersions of a few \kms.

Globular clusters commonly have densities of $\approx 10^3$ \spcc\,
ages of $\simless 10^{10}$ years and velocity dispersions of $\approx
10$ \kms. Precursers of present-day globular clusters, again based
upon expectations from gas expulsion and later evolution, are likely
to have had densities of $\simgreat 10^5$ \spcc\ to $\approx 10^6$
\spcc\ over their first few million years.

\section{Planet formation}

The first potential effect of the stellar environment on any
planetary systems is on their potential for formation. Most
theories for the formation of gas giant planets involve the slow growth
of planetesimals through collisional processes in the
circumstellar disc. This process requires sufficient condensible
material in the disc which in most disc models only exist 
beyond a few \aus from the star. 
In this scenario,
any gas giant closer to the parent star has to undergo an inward migration,
possibly due to the torques from the accretion disc (Lin et. al.~1996).

The growth from planetesimals to a planetary core can take
up to a million years, while the subsequent gas accretion to build a giant 
planet can
take up to 10 million years (eg. Lissauer~1993; Pollack \etal~1996). 
Thus, if extrasolar planet formation occurs in an
analogous fashion as is believed to have occurred in our solar system,
then the primary requirement is that the circumstellar disc is present
and relatively stable over such timescales. In a crowded region, any
stellar encounters can perturb this disc and thus suppress its planet
forming potential. In addition, any encounter that occurs before the
planet has migrated is likely to  eject the planet from the system.

Investigations of stellar encounters including non-self gravitating
circumstellar discs have shown that the encounters generally remove
any material exterior to one-third of the periastron distance (Hall,
Clarke \& Pringle~1996). So, for example, if a star passes within 10
\aus of another, this will remove all disc material exterior  to $\approx 3$
\au.

If we assume that any giant planets will form at separations typical
of the gas giants in the solar system, then any encounters within 10
\aus will suppress the formation of gas-giant planets. Alternatively,
encounters within 50 \aus will truncate the circumstellar disc to
$\approx 15$ \au, which will impede some planet formation and possibly
affect the subsequent migration of any inner planets.  For a given
star in the cluster, the mean time interval, $\tcol$, between
encounters within a distance \rcoll\ is (Binney \& Tremaine~1987):

\begin{equation}
{1 \over \tcol} = 16 \sqrt{\pi} n \vdsp \rc^2 \biggl(1 +{G \ms
\over 2 \vdsp^2 \rc}\biggr),
\label{eqtcoll}
\end{equation}

\begin{figure}
\psfig{{figure=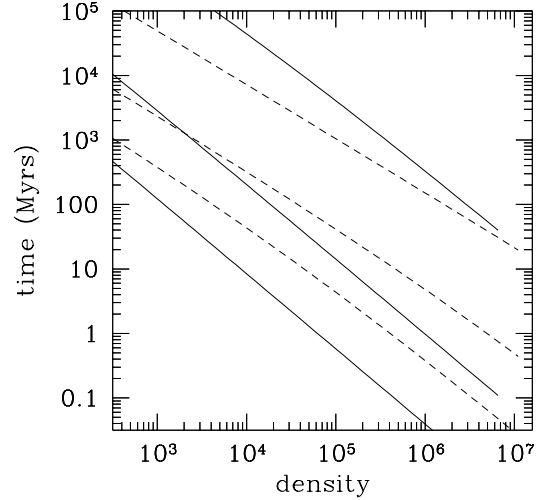,width=3.truein,height=3.truein}}
\caption{\label{detdenstimea}  Encounter timescale versus density for
clusters of 240 (dashed) and 5$\times 10^5$ (solid) stars for (from top to
bottom) 0.5, 10 and 50 \aus separations. The velocity dispersion is taken such that the clusters are virialised.}
\end{figure}

From Equation~\ref{eqtcoll} we can see that the timescale for an
encounter depends on both the stellar density $n$ and the velocity
dispersion $\vdsp$ in addition to the interaction length $\rc$. The
first term in brackets on the right hand side represents the
contribution of the geometric cross section of the target, the second
term represents the effect of gravitational focussing.
Figure~\ref{detdenstimea} plots the timescale for encounters within
0.5, 10 and 50 \aus for clusters containing 240 and $5 \times 10^5$
stars as a function of the stellar density ($\vdsp$ is chosen to
ensure that the clusters are virialised). We see that for encounters
within a disc radius to occur within the planet formation timescales,
$\simless 10^7$ years, requires relatively high stellar densities
($\simgreat 10^4$ \spcc). The required densities are lower for larger
clusters due to their larger velocity dispersions. In order for
encounters to seriously affect planet formation ($\simless 10^7$
years), cluster densities of $\simgreat 10^5$ \spcc\ are
required. Such densities are high relative to today's globular
clusters but are not unreasonable for the earliest phase of a globular
cluster's lifetime before expansion due to gas expulsion.

It is interesting to note that encounters within 0.5 \aus are not
expected to occur on reasonable timescales for most clusters and even
a high density phase would have to be very long ($\simgreat 10^8$
years) in order to have appreciable numbers of encounters that close
to the parent star. Thus, the disruption of close-in gas giant planets
is unlikely to occur due to encounters in a stellar cluster. In this 
context, in order to explain the lack of close-in giant planets
in 47 Tuc (Brown \etal~2000; Gilliland \etal~2000) requires that either the giant planets
did not form or that they did not migrate.

\begin{figure}
\psfig{{figure=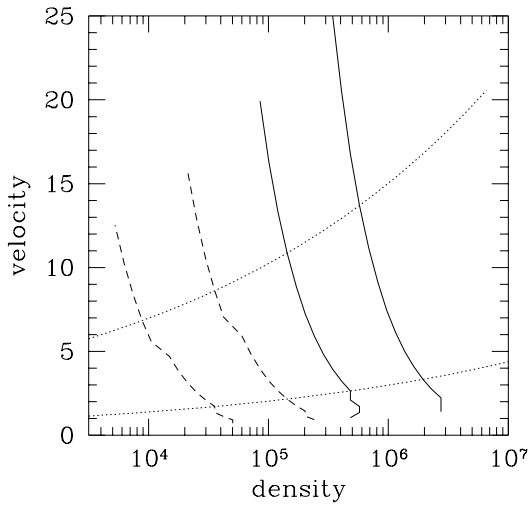,width=3.truein,height=3.truein}}
\caption{\label{destdensvel} 
The cluster conditions in density-\vdisp\ space in order to
have mean encounter timescales of 2 (heavy lines) and 10 (light lines) 
million years for an encounter within 10 \aus (solid lines) and
50 \aus (dashed lines) of a 1.0 \solmas\ star. Probable evolutionary sequences are plotted (dotted
lines) for a cluster of $5 \times 10^5$ and $4 \times 10^3$ stars.  }
\end{figure}

\begin{figure}
\psfig{{figure=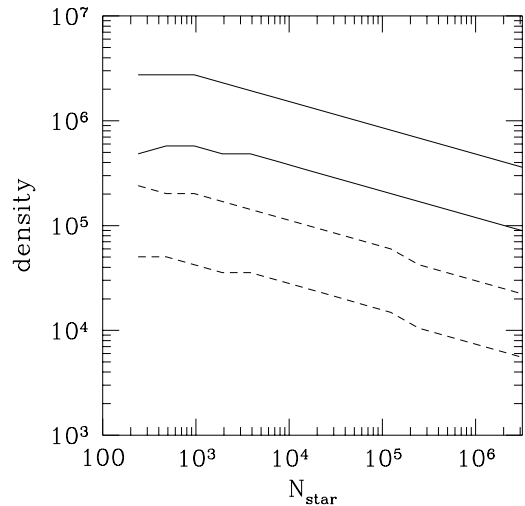,width=3.truein,height=3.truein}}
\caption{\label{destNvel}
Same as Figure~\ref{destdensvel}, but as a function of the 
number of stars and the stellar density in the cluster. }
\end{figure}

Figure~\ref{destdensvel} plots the necessary conditions in terms of
cluster density and velocity dispersions for encounters within 10 and
50 \aus (of a 1.0 \solmas\ star) to occur on timescales of two and ten
million years.  The timescale decreases with increasing density and,
weakly, for increasing velocity dispersion.  For example an encounter
within 10 \aus will occur in ten million years for clusters with
densities $\simgreat 2 \times 10^{5}$ \spcc\ and velocity dispersion
of $\simgreat 10$ \kms\ whereas clusters with densities of $\simgreat
8 \times 10^{5}$ \spcc\ and velocity dispersion of $\simgreat 10$
\kms\ will have similar encounters within one million years.  The
dotted lines in Figure~\ref{destdensvel} show the possible evolutions
in density-velocity space of clusters with $4 \times 10^3$ and $5
\times 10^5$ stars, as they expand (assuming virialised
conditions). Thus, if the planet formation timescale is $\simgreat 5
\times 10^6$ years, stars in a cluster of $3\times 10^5$ stars with
densities $\simgreat 3 \times 10^{5}$ \spcc\ will typically have their
discs stripped through stellar interactions before they are able to
form gas giant planets.

The effect of the number of stars on the necessary cluster conditions
can be seen in Figure~\ref{destNvel} which shows the critical
densities for encounters within 10 and 50 \aus (of a 1.0 \solmas\ star)
on timescales of 2 and 10 million years as a function of the number of
stars in the cluster. The decrease in critical densities in larger-N
systems is due to the larger velocity dispersions in such
systems. From this, it can be seen that densities $\simgreat 10^4$
start to become interesting for disrupting some outer planets or the
outer disc whereas densities $\simgreat 10^5$ are required to affect
planets that form at distances of $\approx 5$ \au.

Combining the expected evolution of clusters of different
numbers of stars and the expected encounter timescale, we expect
that encounters within 10 \aus can occur within the planet forming timescale
in the large-N globular clusters but not in the smaller-N open (embedded) clusters. This difference is mainly due to the expected densities in the earliest
stages of the cluster evolution combined with the higher critical densities
for encounters in the low-N systems.

\section{Planetary system disruption due to stellar encounters}

\begin{figure*}
\psfig{{figure=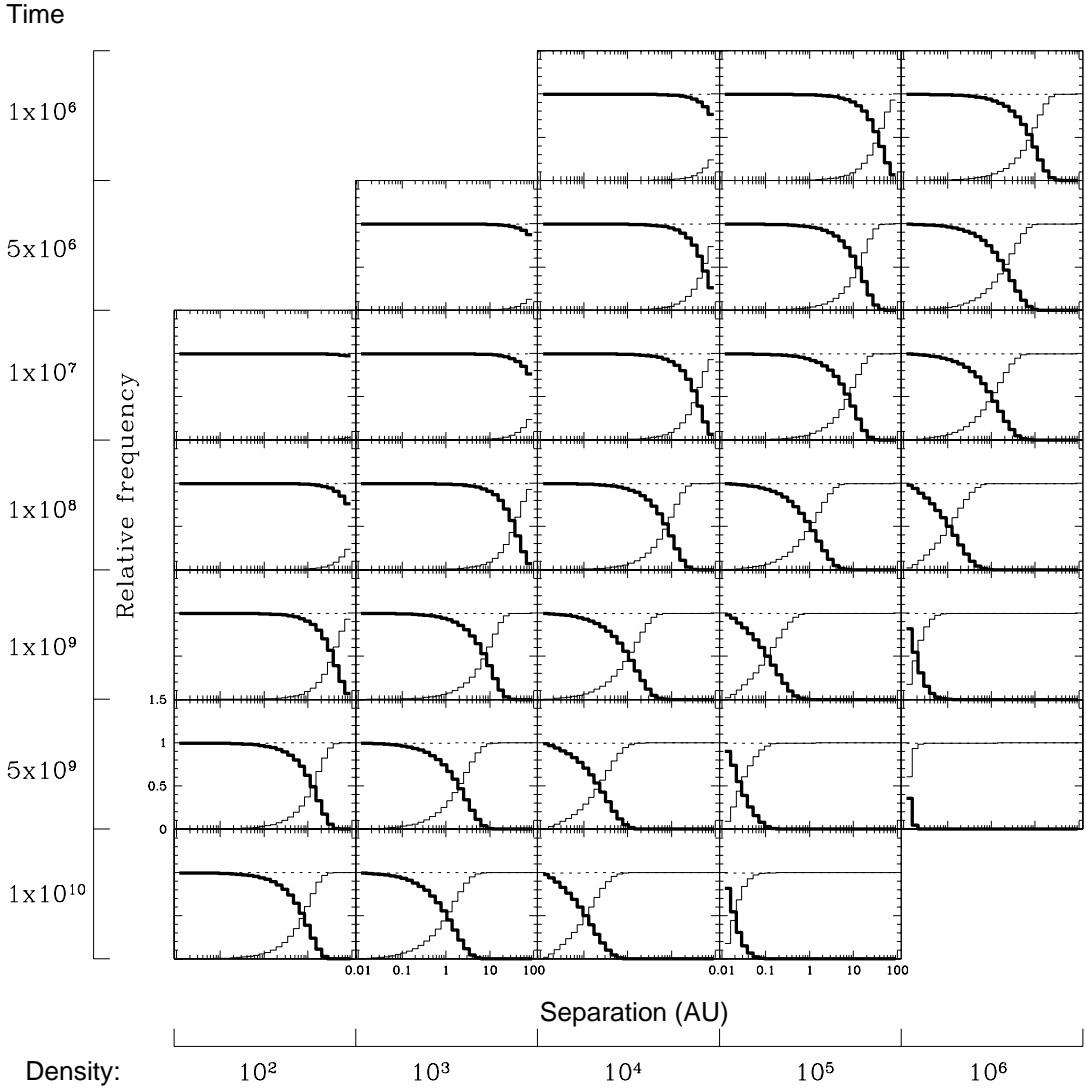,width=6.truein,height=6.truein}}
\caption{\label{case11} Separation distributions for populations of
planetary systems exposed to different globular cluster environments.  In each
case, $M_1=M_3=0.5 \solm$, $M_2=0.001 \solm$ (i.e. one Jupiter
mass).  The density of the cluster varies between $10^2$ to $10^6$ \spcc\ 
 along the x-axis and the total timescale of the
simulation varies from $10^6$ years to $10^{10}$ years down the
y-axis. The dotted line is the initial population while
the final population is the heavy solid line. The lighter solid line
represents the ionised systems. The velocity dispersion is 10 \kms, typical for a globular cluster.}
\end{figure*}

Planetary systems that do succeed in forming in a stellar cluster are then
subject to disruptions from stellar encounters.  Once a planet has
formed, and possibly migrated to its final separation from the parent
star, the probability of a disruptive encounter depends on this
separation and on the cluster properties.  In order to quantify this
probability we performed simulations of the evolution of a population
of planets in various cluster conditions. 

\begin{figure*}
\psfig{{figure=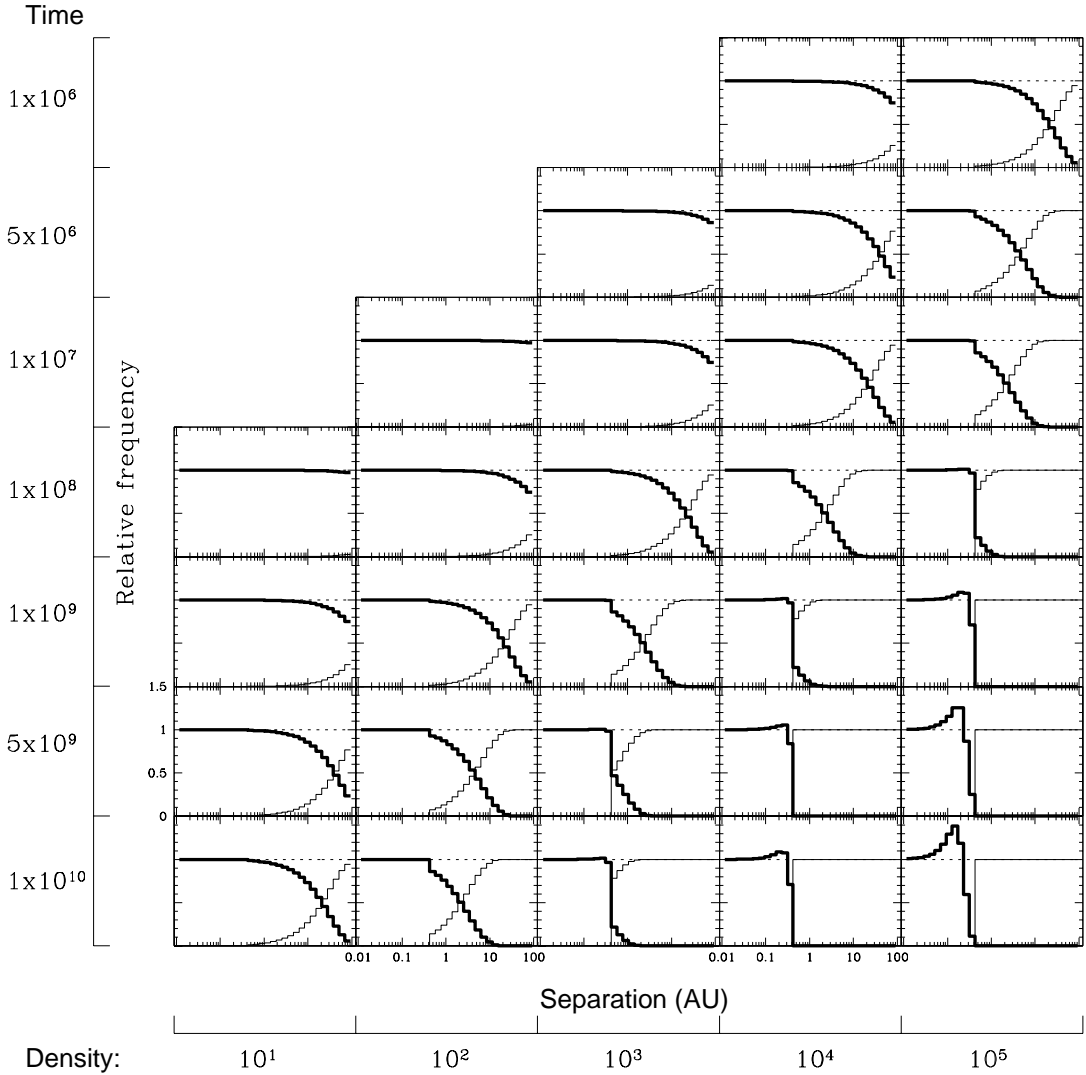,width=6.truein,height=6.truein}}
\caption{\label{case12} Same as Figure~\ref{case11}, but with a velocity
dispersion (of 2 \kms) and densities typical for an open cluster.}
\end{figure*}

The initial distribution
of planetary orbits was taken to be flat in log separation, and spanned a range
of 0.01 to 100 \au. Two main cluster types were investigated, globular
clusters and open clusters. For the globular cluster case, we chose a 
velocity dispersion of 10 \kms. For the open cluster case, the 
velocity dispersion is 2 \kms.
We then used Equation\~(\ref{eqtcoll}) to calculate the probability
of encounters within various time intervals and cluster densities
(see Figures~\ref{case11} and~\ref{case12}). In each case, the total 
time interval was split into smaller time steps, so that multiple
encounters were possible. 

The effect of an encounter on a planetary
system was determined as follows. Where the kinetic energy of the
perturber was greater than the binding energy of the planet, the 
planetary system was assumed to be detroyed (or `ionised'). Where
the kinetic energy of the perturber was less than the binding energy
of the planetary system, the planet was assumed to lose energy
and move closer to its parent star (the system is `hardened'). This
hardening was taken to be 25\%, based on average values obtained
in binary-single star scattering experiments
(Sigurdsson \& Phinney 1993; Davies 1997). In the former case, the 
system is termed `soft', while in the latter, it is said to be 
`hard'. The hard/soft boundary is given by
\begin{equation}
\rhd= \frac{GM_{1}M_{2}\left(M_{1}+M_{2}+M_{3}\right)}{M_{3}\left(M_{1}+M_{2}\right)\venc^{2}},
\label{hardsoft}
\end{equation}
where $M_1, M_2, M_3$ are the masses of the primary, secondary (in
this case planet) and of the perturber star. The encounter velocity,
$\venc$ is essentially the velocity dispersion, \vdisp, of the cluster.

From Equation~\ref{hardsoft} we see that most planetary systems, where
$M_2$ is small, are `soft' and will easily be ionised through
encounters. The question of hardenning is therefore not crucial to
our conclusions. In addition, encounters can drive
eccentricity into the planetary system, and thus potentially further
instability (Davies \& Sigurdsson~2000).

The results from our simulations are divided into two sections
depending on the chosen velocity dispersion. Firstly we present the
results appropriate for Globular clusters with $\vdsp =10$ \kms, then
those for open clusters with $\vdsp = 2$ \kms.

\begin{figure*}
{\centerline{\hbox{\psfig{figure=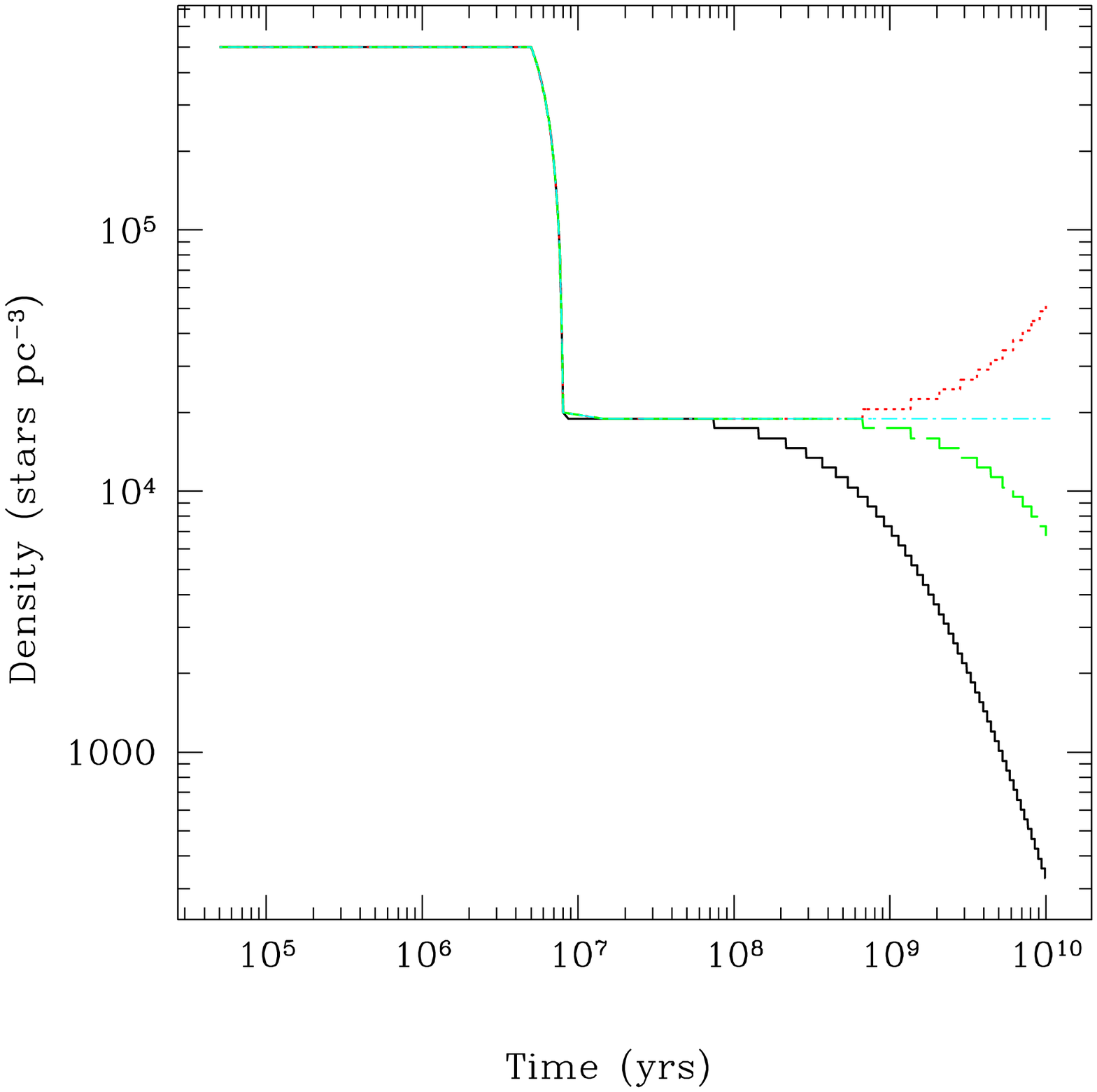,width=3.25truein,height=3.25truein}\psfig{figure=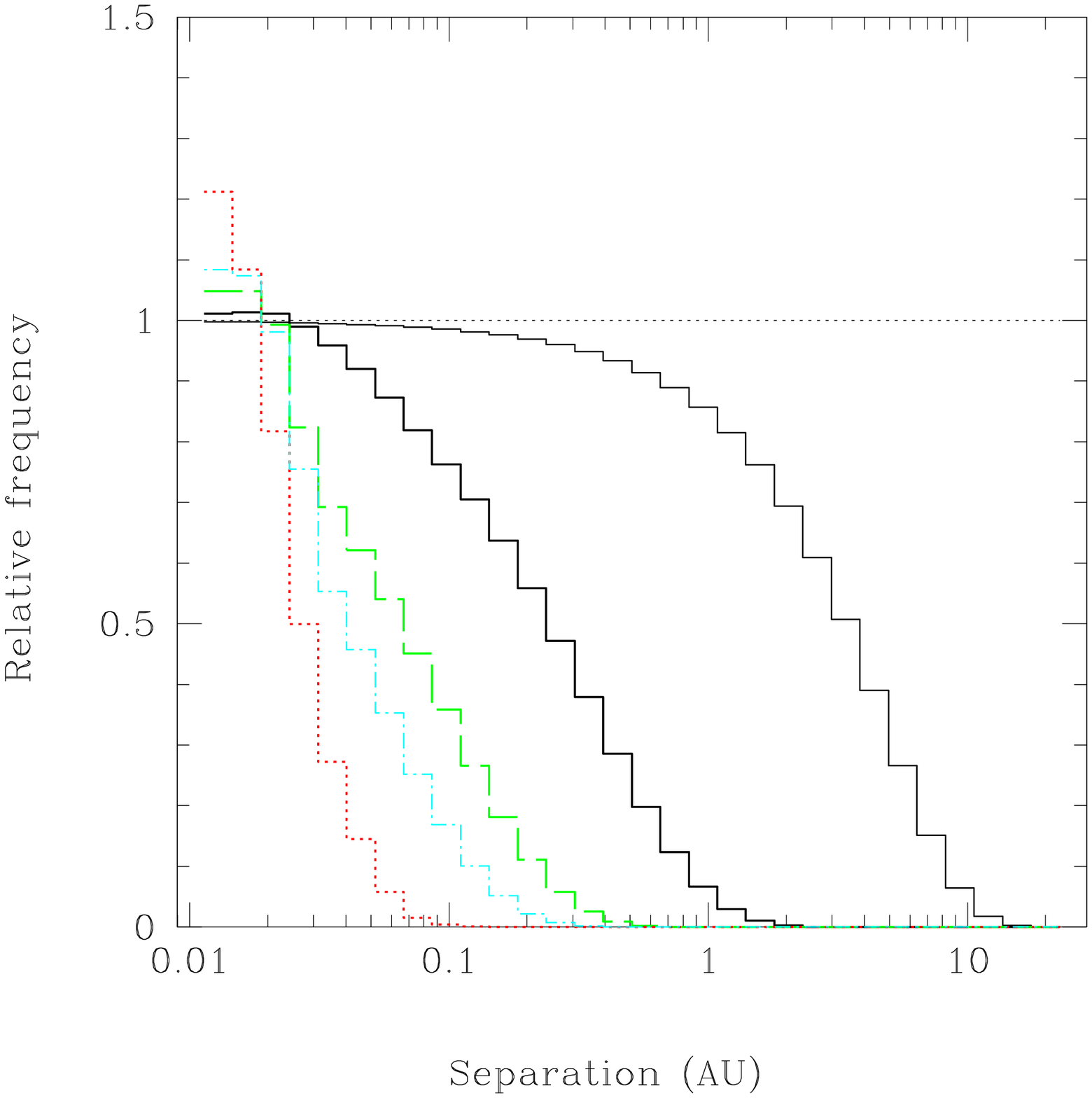,width=3.25truein,height=3.25truein}}}}
\caption{\label{evolclus} The left panel shows the assumed density
evolution in a model cluster containing $5\times 10^5$ stars. The
evolution starts from a density of $5\times 10^5$\spcc\ that expands
after a few $\times 10^6$ years due to gas loss.  Further expansion of
the cluster is taken to occur on one (solid-line), ten (dashed-line)
and $\infty$ (dot-dashed line) \trel. A possible increasing density,
as one might expect as the cluster evolves towards core collapse, is also modelled (dotted line).  The right panel
shows the resultant surviving planetary system distributions after the
initial expansion due to gas loss and after $10^{10}$ years for each
model. They are from left to right the increasing density model
(dotted), the constant density model ($\infty$ \trel, dot-dashed), the
10 \trel evolution (dashed) and the 1 \trel evolution (solid). The
second solid line represents the evolution after the gas-expansion
phase whereas the horizontal dotted line is the initial distribution of
planetary systems.}
\end{figure*}

\subsection{Planetary disruption in Globular clusters}

Figure~\ref{case11} presents the results for the planetary systems in
clusters where the velocity dispersion is $\approx 10$ \kms.  The
figure is broken up into different panels each appropriate for a
specific stellar density and for a specific amount of time. Globular
clusters have typical core densities of $\approx 10^4$\spcc, mean
densities of $\approx 10^3$ \spcc\ and lifetimes that extend to the
age of the Galaxy ($10^9 \simless t \simless 10^{10}$ years).  The cluster
densities increase from left to right and the time increases from top
to bottom. The dotted lines give the initial planetary distribution
while the heavy solid lines give the final planetary distribution. The
light solid lines indicate the fraction of planetary 
systems, and their initial separations, that have been ionised.

The overall result is that the wider (softer) systems are more easily
disrupted than are the tighter systems. Their larger cross section for
an encounter results in lower critical densities and in shorter
encounter timescales. Tighter systems require higher stellar densities
in order to ensure a reasonable timescale for an
encounter. Furthermore, a cluster of given stellar density will
disrupt increasingly tighter systems with time until reaching the
hard/soft boundary. Thus, wide systems are disrupted in most clusters
although they do require longer timescales in the least dense
clusters. In contrast, the tighter systems are only disrupted in
sufficiently old, dense, clusters.

It can be seen that in the case of $\vdsp = 10$ \kms, all systems are
soft. There is no hardening (moving planets to smaller separations)
and any system that is perturbed is disrupted. This occurs as the
hard/soft boundary is at $\rhd \approx 0.02$ \aus or 4 \solrad (smaller
than the separations considered here). Thus some hardening can be
expected of only the tightest of planetary systems.

It should be noted that probable evolutionary paths (based on the
discussion of \S~2) of a globular cluster through this diagram will be
from the upper right (gas-rich cluster) towards the lower left. In
general, any high density phase, will last for $t \simless 10^7$ years
whereas subsequent phases will last successively longer as the cluster
evolves on a relaxation timescale. It is also worth noting that the
long time-periods that a cluster spends in the less dense phases can
actually be more destructive of the relatively short-period planetary
systems than are the short-lived high-density phases.

Present day globular clusters such as 47 Tuc have mean stellar
densities (near the half-mass radius) and ages that put them in the
lower-left part of Figure~\ref{case11} (densities of $\approx 10^3$
\spcc\ and ages of $\simless 10^{10}$ years).  Thus, any planetary
systems with separations $\simgreat 1$ \aus are likely to have suffered
a disruptive encounter and no systems with separations greater than 10
\aus should still exist. In the core of such a cluster with densities
$\simgreat 10^4$ \spcc, even those systems as tight as 0.1 \aus are
likely to be disrupted and only those with separations $\simless 0.01$ \aus
are completely safe.  Even these systems can be disrupted if the core
density is as high as $10^5$ \spcc. Lower densities typical of the
halo of the cluster will leave wider systems ($\simless 10$ \au)
intact.

Planetary systems with very small separations $\simless 1$ \aus are 
generally unaffected by stellar encounters in globular clusters
unless the cluster density was extremely large ($\simgreat 10^5$) for
the majority of the cluster's lifetime ($t \simgreat {\ \rm few\ }
10^9$ years). Thus, the only way to destroy tight planetary systems in
globular clusters is to destroy the disc before the planet forms and
before it has a chance to migrate inwards. If the planet forms in
situ, then stellar encounters are not a promising way of disrupting
the system.

\subsection{Planetary disruption in Open clusters}

Figure~\ref{case12} shows the contrasting case more typical of open
clusters where the velocity dispersion is a few \kms. The first
difference to note is that the hard/soft boundary for planetary systems
is significantly further out ($\approx$ 0.6 \au) and that a cluster with large
densities for long time periods could experience significant
hardening. Typical densities and lifetimes of open clusters ($10^2$
\spcc\ and $t \simless 10^9$ years) precluding significant hardening
of planetary systems as the probability of encounters near the
hard/soft boundary is not very high.  In general, most planetary
systems are not adversely affected by stellar encounters in open
clusters. Significant disruption only occurs for systems with
separations $\simgreat 10$ \au. Even an early high density phase is
unlikely to be sufficiently dense or long-lived to cause much
disruption to any but the widest of planetary systems.

\subsection{ Model Cluster evolutions}

In order to illustrate how the cluster evolution might affect the
planetary populations, we have repeated the simulations of \S 5.1 and 
\S 5.2 but with time-dependent parameters (density and velocity
dispersion), designed to mimic a simple cluster evolution
model. An example of this is included in Figure
\ref{evolclus} for a cluster containing $5 \times 10^5$ stars.  We
assume an initial stellar density of $5 \times 10^5$ \spcc\ during the
gas rich phase which persists for several $\times 10^6$
years.  Once the gas is expelled, the density decreases rapidly to $2 \times
10^4$ \spcc. The subsequent evolution has four possibilities. Firstly
a constant density representing a non-evolving cluster. Secondly a
cluster that continues to expand on its relaxation timescale or,
thirdly, on ten relaxation times. These model evolutions include,
in a heuristic way, the
effects expected from relaxation (Chernoff \& Weinberg~1990) and from
binary heating  (Giersz \& Heggie~1997).  Lastly, a
model where the cluster expands on ten relaxation times is included
to illustrate the effects of core-collapse on any planetary systems in
the cluster core. It should be noted that the different evolutionary models
can be taken to represent different parts of the cluster.

Each model is evolved up to $10^{10}$ years, while the velocity
dispersion is adjusted such that the cluster is virialised. We see that
the initial high density phase can disrupt any systems (or discs) that
extend to several (to ten) \aus or more. Systems with smaller
separations are disrupted during the following less-dense phases due to their
longer timescales. The fastest evolving model destroys the fewest planetary
systems but still leaves only those closer than $\simless 1$ \aus with
a 50 \% chance of survival at $\approx 0.3$ \au. The
model that evolves on ten relaxation times destroys almost all systems $\simgreat 0.2$ \aus whereas the constant density model removes systems $\simgreat 0.1$ \au. The model with increasing density such as occurs in a core-collapse
removes systems $\simgreat 0.05$ \aus with a 50 \% survival rate interior
to $0.03$ \au. It is interesting to note that these last two evolutions
include significant hardening inside $0.02$ \aus ($\approx 4 \solr$). Thus
in this simplified model there should still be some gas giants on very
tight orbits.

\subsection{The Sun's natal environment}

An interesting question to pose is what can we deduce of the probable
natal environment of the Sun and whether such an environment would
leave a detectable trace in the solar system. The existence of the
planets in the solar system and beyond them of the Kuiper belt to
$\approx 50$ \aus (Jewitt \& Luu~2000) implies that any stellar
encounters must have been more distant than that. Thus, it is unlikely
that the Sun spent a significant fraction of its lifetime in a high
density environment.  Alternatively, the apparent lack of Kuiper belt
objects beyond 50 \aus (Hahn~2000) could imply a stellar encounter that
truncated the solar system at that radius (see also Ida, Larwood \&
Burkert~2000).  If this is the correct interpretation, we can deduce
from Figure~\ref{case12} the probable cluster properties in order to
have a significant probability of such an encounter. Thus, the Sun
could have been a member of an open cluster with a mean density of
$\approx 10^2$ \spcc\ for $t > 10^8$ to nearly $10^9$ years, or
alternatively, that the cluster had an earlier high-density phase with
either a density of $\approx 10^3$ \spcc\ for $t > 10^7$ to nearly
$10^8$ years or a density of $\approx 10^4$ \spcc\ for $t\approx$ a
few $\times 10^6$ years. The existence of the Oort cloud beyond the
limit of the Kuiper belt implies that any interaction would have had
to happen before the bulk of the Oort cloud was ejected from the solar
nebular disc, thus within a few $10^7$ years. This is possible if the
sun was born in a cluster with a density of $n \simgreat 10^{3}$ \spcc
that dissolved, or evolved to a lower density, within $\approx 10^8$
years.

\section{Free-floating planets}

One of the implications of stellar encounters disrupting planetary
systems in stellar clusters is that there should then be a population
of free-floating planets. Such a population due to stellar encounters
is unlikely to be significant in most open clusters as these clusters
are not sufficiently long-lived.  Even fewer free-floating planets are
expected in the young clusters due to the disruption of planetary
systems.  Alternatively, the more frequent disruption of planetary
systems in globular clusters should result in a population of
free-floating planets. For example, in clusters with ages $\approx
10^{10}$ years and densities $\simgreat 10^3$ \spcc, any planetary
systems with separations $\simgreat 1$ \aus should have been disrupted
and these planets liberated into the cluster. If such systems resemble
our own, then a number of planets would be liberated per event. The
likelihood for finding free-floating planets depends on their
velocities.  In open clusters any free-floating planets are likely to
have velocities after disruption well in excess of the escape speed
(Smith \& Bonnell~2000).  In contrast, post-disruption velocities in
Globulars are more likely to be comparable or less than the higher
escape speeds there (Smith \& Bonnell~2000). In either case, due to
their low-mass, subsequent two-body encounters will increase their
velocity dispersion and thus limit their lifetime in the cluster.

\section{Conclusions}

We have shown that young planetary systems similar to the Solar system
have a good chance of surviving their early years if they have the
good fortune to form in an open cluster environment. Planetary systems
formed in globular clusters, on the other hand, face two major
obstacles to their reaching adulthood. Firstly, the natal disk from
which planets could form must survive for at least a few million
years. If the earliest stages of a globular cluster include a high-density
phase ($\simgreat {\rm few } 10^5$ \spcc) 
then the circumstellar disc can be truncated inside the region
where gas-giant planets are believed to form.  
Such a high density phase
is consistent with expectations based on the efficiency of star formation
in nearby embedded (young) stellar clusters and on subsequent cluster
dynamics. This could explain the recent find of a lack of close planets
in 47 Tuc (Brown \etal~2000, Gilliland \etal~2000).

Secondly, the planetary
systems must endure a constant bombardment from neighbouring star 
systems. These interactions will destroy most planetary systems 
beyond about 0.3 \aus over the lifetime of the cluster. We conclude that
giant planet formation may be made more difficult in a globular
cluster and that any that do form are unlikely to 
survive unless their orbits are $\simless 0.3$ \au. 

The planets thrown out by interactions would be expected to form 
a population of free substellar bodies in the cluster. The velocity 
obtained by them from the initial disruption and subsequent encounters 
are likely to be higher than the escape velocity of the cluster
and thus free-floating planets should not form a significant population
in stellar clusters.

\section{Acknowledgements}
We thank Simon Goodwin, Pavel Kroupa and Douglas Heggie for insightful
discussions on the cluster evolution. We also thank the referee, John
Chambers, for his suggestions.  for IAB acknowledges support from a
PPARC Advanced Fellowship. MBD gratefully acknowledges the support of
the Royal Society through a URF.

\label{lastpage}

\end{document}